\documentclass[conference]{IEEEtran}

\usepackage{hyperref}
\usepackage{tabu}
\usepackage{graphicx,epsfig,color}
\usepackage{tikz}
\usepackage{pgfplots}
\pgfplotsset{compat=1.14}
\usepackage{csquotes}
\usepackage{cite}
\usepackage{subfig}
\usepackage{amsfonts,amssymb,amsmath,bm}
\begin{document}
%
\title{Methodology for Multi-stage, Operations- and Uncertainty-Aware Placement and Sizing of FACTS Devices in a Large Power Transmission System}

\author{{\bf Vladimir~Frolov} $^{(a)}$
        and {\bf Michael~Chertkov} \textit{Senior Member, IEEE} $^{(a,b)}$ \\
        $^{(a)}$ Skolkovo Institute of Science and Technology, Nobel Street 3, Moscow Region, 143026, Russia\\
$^{(b)}$ Center for Nonlinear Studies and Theoretical Division, T-4, LANL, Los Alamos, NM 87545, USA\\
E-mails: vladimir.frolov@skolkovotech.ru \& chertkov@lanl.gov
}


%


\newcommand{\norm}[1]{\left\lVert#1\right\rVert}
\newcommand{\myspecial}[1]{\mathrm{#1}}

\maketitle

\begin{abstract}
We develop new optimization methodology for planning installation of Flexible Alternating Current Transmission System (FACTS) devices of the parallel and shunt types into large power transmission systems, which allows to delay or avoid installations of generally much more expensive power lines. Our methodology takes as an input projected economic development, expressed through a paced growth of the system loads, as well as uncertainties, expressed through multiple scenarios of the growth. We price new devices according to their capacities. Installation cost contributes to the optimization objective in combination with the cost of operations integrated over time and averaged over the scenarios. The multi-stage (-time-frame) optimization aims to achieve a gradual distribution of new resources in space and time. Constraints on the investment budget, or equivalently constraint on building capacity, is introduced at each time frame.  
Our approach adjusts operationally not only newly installed FACTS devices but also other already existing flexible degrees of freedom. This complex optimization problem is stated using the most general AC Power Flows.  Non-linear, non-convex, multiple-scenario and multi-time-frame optimization is resolved via efficient heuristics, consisting of a sequence of alternating Linear Programmings or Quadratic Programmings (depending on the operational cost dependence on the power injected by the generators) and AC-PF solution steps designed to maintain operational feasibility for all scenarios. Computational scalability and other benefits of the newly developed approach are illustrated on the example of the 2736-nodes large Polish system. 
One most important advantage of the framework is that the optimal capacity of FACTS is build up gradually at each time frame in a limited number of locations, thus allowing to prepare the system better for possible congestion due to future economic and other uncertainties.
\end{abstract}


%
\IEEEpeerreviewmaketitle

\section*{Nomenclature}
\begin{IEEEdescription}[\IEEEusemathlabelsep\IEEEsetlabelwidth{$\overline{P}_g$ ($\underline{P}_g$) $\in$ $\mathbb{R}^{N_g}$}]
\item [\underline{Parameters}:]
\item
\item[$N_l$] Number of power lines in operation 
\item[$N_b$] Number of buses in the system 
\item[$M$] Number of loading scenarios representing given time frame
\item[$N$] Number of scenarios representing planning horizon
\item[$T$] Number of time frames representing horizon
\item[$t=1..T$] Index of a decision point
\item[$a=1..M$] Index of a scenario at time frame t
\item[$Pr_{t,a}$] Occurrence probability of a scenario $a$ at time frame $t$
\item[$x_0$  $\in$ $\mathbb{R}^{N_l}$] Vector of initial line inductances 
\item[$\overline{P}_G$ ($\underline{P}_G$) $\in$ $\mathbb{R}^{N_b}$] Vector of maximum (minimum) active power generator outputs 
\item[$\overline{Q}_G$ ($\underline{Q}_G$) $\in$ $\mathbb{R}^{N_b}$] Vector of maximum (minimum) reactive power generator outputs 
\item[$P_{D_0}$ ($Q_{D_0}$) $\in$ $\mathbb{R}^{N_b}$] Vector of active (reactive) power demands 
\item[$\overline{S}$ $\in$ $\mathbb{R}^{2 N_l}$] Vector of line apparent power limits 
\item[$\overline{V}$ ($\underline{V}$) $\in$ $\mathbb{R}^{N_b}$] Vector of maximum (minimum) allowed voltages 
\item[$C_{SC}$ $\in$ $\mathbb{R}$] Cost per Ohm of a series FACTS device 
\item[$C_{SVC}$ $\in$ $\mathbb{R}$] Cost per MVAr of a shunt FACTS device 
\item[$N_{years}$ $\in$ $\mathbb{R}$] Planning horizon 
\item
\item [\underline{Optimization variables (operational, scenario dependent):}]
\item
\item[$V$ $\in$ $\mathbb{R}^{N_b}$] Vector of bus voltage magnitudes
\item[$\theta$ $\in$ $\mathbb{R}^{N_b}$] Vector of bus voltage angles
\item[$P_G$ $\in$ $\mathbb{R}^{N_b}$] Vector of generator active power injections
\item[$Q_G$ $\in$ $\mathbb{R}^{N_b}$] Vector of generator reactive power injections
\item[$x$ $\in$ $\mathbb{R}^{N_l}$] Vector of line inductances modified by SC devices
\item[$\Delta x$ $\in$ $\mathbb{R}^{N_l}$] Vector of series FACTS settings
\item[$\Delta Q$ $\in$ $\mathbb{R}^{N_b}$] Vector of shunt FACTS settings
\item [\underline{Optimization variables (investment, scenario independent):}]
\item
\item[$\overline{\Delta x}^t$ $\in$ $\mathbb{R}^{N_l}$] Vector of series FACTS capacities built at decision point t
\item[$\overline{\Delta Q}^t$ $\in$ $\mathbb{R}^{N_b}$] Vector of shunt FACTS capacities built at decision point t
\end{IEEEdescription}

\IEEEpeerreviewmaketitle

\section{Introduction}
%
%
%
%


\IEEEPARstart{E}{nergy} market deregulation and massive installation of renewables are among significant stress factors forcing transmission systems across the world to operate closer to their limits. 
When a transmission system is constrained building new lines seems a natural remedy
 \cite{03AMC,12BTB,ref1,LANL_Resilient_Grid_Design}. However, this option is costly and severely limited in many countries due to social and environmental concerns, hence rising the question: if the system can be upgraded creatively and gracefully by installing Flexible Alternating Current Transmission System (FACTS) devices which are both less expensive and whose spatial footprint is much smaller?

This question was addressed in the literature  \cite{ref2,ref3,ref4,ref5,ref6} suggesting
that FACTS devices may indeed be effective in increasing transmission capacity  and keeping the system safe and operational as the demand growth. However, it also became clear from the studies that this sustainable upgrade needs to be creative -- types, locations and capacities of the newly introduced FACTS devices must be chosen carefully in order to exploit their benefits. Several objectives, including decreasing operational \cite{ref4,ref8,ref19} and investment costs \cite{ref2,ref6}, reducing transmission losses \cite{ref3,ref8}, increasing power system loadability and managing system congestion \cite{ref7,ref8,ref11,ref18}, reducing load curtailment \cite{ref12} and improving voltage profile \cite{ref6} and voltage stability index \cite{ref5,ref18}, have been considered. Formally, these problems were stated as mixed Integer Nonlinear Programming (MINLP) optimizations which exact solutions are limited to only very small (typically not exceeding tens of nodes) power system models. Sensitivity analysis \cite{ref12,ref13}, Mixed Integer Linear Programming (MILP) \cite{ref16} and genetic algorithms \cite{ref9,ref10,ref14,ref21} were suggested to resolve the MINLP formulations approximately.
The sensitivity based methods consist in computing a few indicators to identify the most critical lines which are thus suggested as good option for FACTS placement. However,  this ad-hoc methodology is obviously not optimal and it also does not offer any suggestion on how to size the devices. Genetic algorithms aim at finding optimal solutions but come with unacceptably high computational cost. Relaxation and approximation techniques were suggested to convert MINLP to MILP \cite{ref17}. Approximation techniques are computationally less expensive as they use the simplified line flow based model \cite{ref6} or DC power flow model \cite{ref16}. However, such approximations have serious limitations especially for planning installation of shunt FACTS devices.

One significant complication in resolving MINLP is related to combinatorial explosion in the number of choices for possible FACTS placements. It is thus desirable to enforce sparcity of the installation. A principal way to achieve sparse placement in a computationally feasible way was suggested in \cite{ref22,ref23}. An alternative way to achieve sparse placement was also discussed in \cite{ref17}.

Uncertainty of the economic growth is another notable obstacle for efficient and practical implementation of this and other related planning problems. A solution for resolving this problem was suggested \cite{16FGBBC} where  the uncertainty was modeled via exogenously described multiple operational samples of loads and related nonlinear AC power flow solutions was embedded in the large-scale optimization explicitly.

In this manuscript we extend the approach of \cite{16FGBBC} to resolve the last remaining issue in the optimal multiple-scenario aware, AC-based and sparse placement and sizing of the FACTS devices in a large transmission system.  Here, we choose to represent future not in one time step, as was done in \cite{16FGBBC}, but in multiple time steps. In other words here we complete general description of the framework started in \cite{ref22,ref23,16FGBBC} and propose a comprehensive resolution for finding optimal locations of FACTS devices in a large transmission system by preparing the system for future loading gradually through multi-stage, properly paced investments. Main highlights of our comprehensive approach are as follows:


\begin{enumerate}
\item Planning horizon is represented by multiple decision points (multiple time frames). At each new time frame a set of new FACTS devices can be installed, and they are assumed available for operations immediately such that respective operational values do not exceed the installed capacities.  Therefore, installation of FACTS devices can be paced. 
In this manuscript we work with a finite number, $T$, of the time-horizon sub-intervals. Notice, however, that extension of the approach to the case of a receding horizon, thus accounting at each step for updated forecast,  is straightforward. (We plan to conduct extensive discussion of the  receding horizon experiments in a future journal version of the manuscript.) 
    
\item Future operational conditions are represented through multiple loading scenarios and associated probabilities broken into time frames. Our framework is set in the way that the scenarios are stated as an exogeneous input,  which allows us to separate the problem of scenario generation from the intrinsic optimization details. 
Given the exogeneously prescribed scenarios, optimal installation of FACTS is resolved within the optimization framework by accounting for both investment variables and operational variables, characterizing  installation decisions and operational implementations (per scenario) respectively. It is important to stress that an optimal constraints (i.e. feasible) all the scenarios. (This is in a contrast with the worst case planning approach.) 

\item Both capital and operational expenditures are optimized simultaneously. To the best of our knowledge no prior works have considered optimizing them at the same time. But for practical planning horizons operational cost is much bigger than the cost of FACTS installation. Thus relatively small additional investment allows to save a significant amount of money by reducing congestion additionally to resolving infeasibility of particular load scenarios.

\item A novel optimization iterative heuristics is developed, which is a combination of analytic linearization of non-linear constraints, a solution of Quadratic Programming (QP) or Linear Programming (LP) (depending on generation cost) problem for finding of investment variables and operational settings for all scenarios and Alternating Current (AC) Power Flow (PF) resolution (for each scenario) to update previously found states.

\item The developed heuristics for finding optimal locations of FACTS devices considering multiple loading scenarios and multiple time frames can be applied to large power systems. In other words, the developed approach is scalable. To the best of our knowledge the system considered in the literature addressing optimal locations of FACTS placement consists of a maximum of 1228 buses \cite{ref7}. The results of the proposed methodology is demonstrated on 2736-bus Polish system, thereby proving its scalability. Moreover, algorithm provides upper bound solutions of the objective function with the gap less than 0.1$\%$
\end{enumerate}




The material in the manuscript is organized as follows. Our basic optimization model is introduced in Section \ref{sec:Opt}. The solution algorithm is described in Section \ref{sec:algorithm}. Section \ref{sec:case_studies} is at the core of the manuscript - it describes in details our experiments.  We conclude and discuss path forward in Section \ref{sec:conclusions}. Appendixes describe constructions and steps needed to run our algorithm. (This material is largely a repetition from \cite{16FGBBC} reproduced here for completeness.)

\section{Optimization model}
\label{sec:Opt}

This section describes our optimization framework for operations-aware installation of FACTS devices taking into account multiple future decision points (or multiple time intervals). 

Assume that the planning time horizon is  $N_{years}$, $T$ is the number of time intervals, $M$ is number of given loading configurations (scenarios) per each time frame (the number of scenarios per time frame may also vary with the time frame). In this setting we aim to place and size the Series Compensation (SC) and Static Var Compensation (SVC) devices, where an SC device, installed at a line,  modifies inductance of the line (thus allowing to reroute apparent power),  while an SVC device, installed at a node, injects or consumers reactive power at the node thus helping to balance the voltage locally. 

\begin{figure}[h!]
\centering
\includegraphics[width=0.5\textwidth]{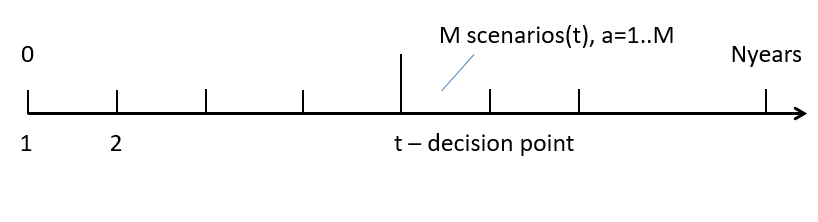}
\caption{Illustration of the relation between the number of scenarios, $M(t)$ (each defined in the time interval, $t$) and the number of time intervals, $T$. $N=\sum_{t=1}^T M(t)$ is the total number of constraints imposed (per line or per node) within our optimization formulation. (See text for additional explanations.)
\label{Mframes}}
\end{figure}

Fig.~\ref{Mframes} illustrates the setting. Since scenarios are generated within each time interval independently, the total number of paths accounted for within our optimization formulation is $\prod_{t=1}^T M(t)$, where a path is a sequence of $T$ scenarios (each per time interval). Notice that even though the number of paths is exponential in $T$, the total number of the operational constraints in the optimization formulation, $N=\sum_{t=1}^T M(t)$, scales linearly in $T$.

The overall problem is to minimize a combination of the sum (over the time intervals) of the investment cost and the sum of operational costs over all the scenarios taking into account (a) operational constraints for every scenarios (per time interval) and (b)  investment constraints requiring that the operational variables (for every scenario per tiem interval) do not exceed the respective installed capacities.  {\bf Operational settings can be different for different scenarios but installed capacities of the devices are the same for all the scenarios representing given time interval}. 

Mathematically the optimization problem is stated as follows:
\begin{eqnarray}
 \min_{\overline{\triangle x}, \overline{\triangle Q}, y^{(a)}_{t}}  C_{SC}\!\  \sum_{t=1}^{T}\sum_{\{i,j\}\in{\cal E}}\overline{\Delta x}_{ij}^{t} + C_{SVC}\sum_{t=1}^{T}\sum_{i\in{\cal V}_l}\overline{\Delta Q}_{i}^{t}\nonumber\\
+ 8760N_{years}\sum_{t=1}^{T}\sum_{a=1..M}Pr_{t,a}*C_{t,a}(P^{(t,a)})	\label{OPT}\nonumber\\
\end{eqnarray}
subject to:
\begin{align}
& y^{(t,a)} = (x,V, \theta, P, Q)^{(t,a)}	&	\forall a, \forall t	\label{c20}	\\
& \overline{\Delta x}^{t}_{total}=\sum_{h=1}^{t}\overline{\Delta x}^{h} & \forall t	\label{xtot}\\
& \overline{\Delta Q}^{t}_{total}=\sum_{h=1}^{t}\overline{\Delta Q}^{h}	& \forall t \label{Qtot}\\
& \overline{\Delta x}^{t} \geq 0 & \forall t	\label{xtotnz}\\
& \overline{\Delta Q}^{t} \geq 0 & \forall t	\label{Qtotnz}\\
& x^{(t,a)}=x_0^{(t,a)}+\triangle x^{(t,a)} & \forall a, \forall t	\label{c2} \\
& P_G^{(t,a)} = P_{D_0}^{(t,a)} + P^{(t,a)}	& \forall a, \forall t	\label{ac}	\\
& Q_G^{(t,a)} = Q_{D_0}^{(t,a)} + Q^{(t,a)} + \Delta Q^{(t,a)}	& \forall a, \forall t		\label{reac} \\
& P_i^{(t,a)} = \sum_{j\sim i}{\Re(S_{ij}^{(t,a)})} & \forall i, a, t	\label{acinj}		\\
& Q_i^{(t,a)} = \sum_{j\sim i}{\Im(S_{ij}^{(t,a)})}&	\forall i, a, t	\label{reacinj}		\\
& \underline{P}_G^{(t,a)}  \leq P_G^{(t,a)} \leq \overline{P}_G^{(t,a)}	&	\forall a, \forall t \label{c8}\\
& \underline{Q}_G^{(t,a)}  \leq Q_G^{(t,a)} \leq \overline{Q}_G^{(t,a)}	&	\forall a, \forall t \label{cc8}\\
& -\overline{\triangle x}^{t}_{total} \leq \triangle x^{(t,a)} \leq \overline{\triangle x}^{t}_{total} &	\forall a, \forall t \label{c4}\\
& -\overline{\triangle Q}^{t}_{total} \leq \triangle Q^{(t,a)} \leq \overline{\triangle Q}^{t}_{total} & \forall a, \forall t \label{c5}\\
& \underline{V}^{(t,a)} \leq V^{(t,a)} \leq \overline{V}^{(t,a)}	&	\forall a, \forall t \label{c6}
\end{align}
\begin{align}
& [\Re(S)^{(t,a)}]^{T}[\Re(S)^{(t,a)}]+[\Im(S)^{(t,a)}]^{T}[\Im(S)^{(t,a)}]& \nonumber\\
& \leq (\overline{S}^{(t,a)})^2 & \forall a, \forall t \label{c9}\\
&  C_{SC}\sum_{\{i,j\}\in{\cal E}}\overline{\Delta x}_{ij}^{t} + C_{SVC}\sum_{i\in{\cal V}_l}\overline{\Delta Q}_{i}^{t} \leq MaxB^{t} & \forall t \label{maxb}\\
&   \overline{\Delta x}_{ij}^{t} \leq \overline{\Delta x}_{ij}^{t-max}; \overline{\Delta Q}_{i}^{t} \leq \overline{\Delta Q}_{i}^{t-max}   & \forall t \label{maxSC_SVC}
\end{align}
where $a=1,\cdots,M$ labels the scenarios; upper index $t$ labels the time intervals, $t=1,\cdots,T$; ${\cal V}$ and ${\cal E}$ denotes the set of nodes and the set of (undirected) edges, of the grid-graph, where a node can be of  the load type, $i\in{\cal V}_l$, or of the generator type, $i\in{\cal V}_g$.

The objective function in \eqref{OPT} consists of three terms. The first two, sparsity promoting terms \cite{ref22,ref23}, express the capital investment costs of the installation of the two types of FACTS devices (investment can be performed at each decision point $t$). The third term stands for the operational cost in which the summation is over all the scenarios for each time frame and over the time frames ($\forall a$ is a shortcut for, $\forall a=1,\cdots,N$) accounting for respective occurrence probability multiplied by the number of years (service period). Therefore, the optimization \eqref{OPT} is nothing but an operational aware planning.

Each scenario in \eqref{OPT} is stated in terms of the set of operational variables. Description of the optimization constraints in \eqref{OPT} is as follows. \eqref{xtot} and \eqref{Qtot} represent total available capacity at the decision moment $t$. \eqref{xtotnz} and \eqref{Qtotnz} ensures that the already installed capacities are inherited in the future time frames.  \eqref{c2} bounds actual line inductances, which are adjusted according to the operational value of the installed series compensation for each scenario, within their respective installed capacities (represented by \eqref{c4}). \eqref{ac} and \eqref{reac} represent active and reactive power balances at each bus of the network. Components of the vectors $P_G$ ($Q_G$) and $P_{D_0}$ ($Q_{D_0}$) are assumed equal to zero at the  buses containing, respectively, no generators or loads. \eqref{acinj} and \eqref{reacinj} represent the net active ($P \in \mathbb{R}^{N_b}$) and reactive ($Q \in \mathbb{R}^{N_b}$) power injections at the system buses. The term $\Delta Q$ expresses shunt compensation by SVC adjusted to a scenario bounded by the respective installed capacities (represented by \eqref{c5}). Active and reactive power generation limits are set by \eqref{c8} and \eqref{cc8}. Voltage and thermal line flow constraints are represented by \eqref{c6} and \eqref{c9}. $S_{ij}^{(a)}=S_f^{(a)}$ and $S_{ij}^{(a)}=S_t^{(a)}$ stand, respectively, for the apparent power flows from $i$ to $j$ and to  $j$ from $i$ along the line $\{i,j\}$. One also accounts for the budget constraint per time step, \eqref{maxb}, and/or for the maximum built capacity constraint per time step, \eqref{maxSC_SVC}. 

The main challenge in resolving the optimization is related to nonlinearity of the Power Flow relations \eqref{acinj}, \eqref{reacinj} and also to nonlinearity of the line thermal limits \eqref{c9}. Available non-linear solvers, such as IPOPT, are not effective in resolving the nonlinearities for large systems efficiently. This has motivated us to develop an heuristic algorithm consisting in sequential linearization of the nonlinear constrains discussed in the following Section (and,  specifically, in Subsection \ref{subsec:linearization}.

\section{The Algorithm}
\label{sec:algorithm}

This Section describes the algorithm which allows us to resolve efficiently (and in spite of its complexity) the optimization problem just stated. Our algorithm consists of the following steps:
\begin{enumerate}
\item Scenarios are generated for each time frame according to the methodology suggested and described in details in \cite{16FGBBC}. Briefly,  one picks the base case, re-scale it for different time frame (taking into account the economic growth), and then introduce fluctuations around the re-scaled solutions to represent the forecasted load uncertainty. The fluctuations are chosen to be Gaussian with the standard deviation proportional to the mean. 
\item Generation is initialized (for each load scenario) according to scheme explained in Appendix \ref{sec:init_gen}.
\item If some of the constraints \eqref{c2}-\eqref{c9} are violated the initial state of the system is outside of the feasible domain defined by them. The non-linear constraints \eqref{acinj}, \eqref{reacinj} and \eqref{c9} are linearized around the current state. This allows to construct current linearized version of the non-linear optimization problem \eqref{OPT}-\eqref{c9}.
\item The resulting linearized problem is solved by QP (or LP, depends on generation cost functions) using one of the available algorithms of the CPLEX solver \cite{CPLEX}.
\item AC power flow (AC-PF) is solved to update the state obtained at the previous step.  This step is needed to prepare a feasible solution for the next iteration. 
\item Steps 2-5 are repeated till either no constraints remain violated or the target precision is reached or the maximum allowed number of iterations is reached.
\end{enumerate}
It is important to emphasize that, by construction, the algorithm maintains a feasible physical states at each iteration of its main loop including linearization, solution of the current QP optimization and back projection to the non-linear PF equations (achieved through the AC PF step).

Below we present details of the main steps of the algorithm.

\subsection{Linearization}
\label{subsec:linearization}

The Power Flow (PF) equality constraints \eqref{acinj}, \eqref{reacinj} and the thermal limit constraints \eqref{c9} are nonlinear. We choose to add to the optimization formulation auxiliary variables -- active and reactive power flows expressed via voltages, phases and system parameters (see Appendix \ref{sec:system_modeling} for modeling details). This allows to localize non-linearities in local relations between the line power flows and respective voltages and phases at the two ends of the lines. Details of this technical trick, aimed at improving performance of the CPLEX solver, are as follows. 

Introduce the following operational state.
\begin{equation}
y^{(a)} = (x,V,\theta,P,Q,p^{from},p^{to},q^{from},q^{to})^{(a)},
\end{equation}
and substitute the left hand side of Eq.~\eqref{c9} by its Taylor expansion around the current state, $y_{pre}^{(a)}$,
\begin{equation}
F_{pre}^{(a)} + \nabla F^{(a)} (y^{(a)} - y_{pre}^{(a)}) \leq (\overline{S}^{(a)})^2,
\end{equation}
thus arriving at the two equations representing a line
\begin{eqnarray}
&& (p^{from}_{pre})^2+(q^{from}_{pre})^2+2*p^{from}_{pre}*(p^{from}-p^{from}_{pre})+\nonumber\\
&& +2*q^{from}_{pre}*(q^{from}-q^{from}_{pre}) \leq (\overline{S}^{(a)})^2,\nonumber\\ &&
(p^{to}_{pre})^2+(q^{to}_{pre})^2+2*p^{to}_{pre}*(p^{to}-p^{to}_{pre})+\nonumber\\
&& +2*q^{to}_{pre}*(q^{to}-q^{to}_{pre}) \leq (\overline{S}^{(a)})^2, \nonumber
\end{eqnarray}
where the newly introduced auxiliary variables should also be substituted by the respective linearized expressions
\begin{align}
&p^{from}=p^{from}_{pre}+\nabla (p^{from})^{(a)}_{pre} (y^{(a)}-y_{pre}^{(a)})	\nonumber\\
&p^{to}=p^{to}_{pre}+\nabla (p^{to})^{(a)}_{pre} (y^{(a)}-y_{pre}^{(a)}) \nonumber \\
&q^{from}=q^{from}_{pre}+\nabla (q^{from})^{(a)}_{pre} (y^{(a)}-y_{pre}^{(a)}) \nonumber \\
&q^{to}=q^{to}_{pre}+\nabla (q^{to})^{(a)}_{pre} (y^{(a)}-y_{pre}^{(a)})\nonumber
\end{align}

Power balance constraints \eqref{ac}, \eqref{reac}, \eqref{acinj}, \eqref{reacinj} will be exact linear in terms of auxiliary variables.

Three comments/clarifications are in order.  First, notice that the operational variables are adjusted independently for each scenario, thus enabling devices' efficient utilization. Second,  to manage possible degeneracy of the resulting system of linear constraints and limit the change of reactive flows at every we 
add to Eqs.~\eqref{OPT}-\eqref{maxSC_SVC} the following soft constraints for reactive power dispatch at each QP/LP step of the procedure
\begin{equation}
|Q_G^{(a)} - Q_{G_{pre}}^{(a)} | \leq \epsilon.
\nonumber 
\end{equation}
(In the case of LP,  when the degeneracy is stronger, we also add similar constraints imposed on the active power.)
Finally, third, to speed up the QP/LP computations one uses a cutting plane (constraint management) procedure. We split the whole set of constraints \eqref{c9} into ``active" and ``inactive" sets including the constraints which were overloaded and, respectively, not overloaded, at the current state (of the previous iteration) or at any of the preceding steps.  Only active constraints are explicitly accounted for in the optimization, while the validity of the inactive constraints is verified post-factum and the active/passive split is updated at every LP/QP step.



\subsection{QP/LP implementation}
Standard CPLEX solver is called at each QP/LP step which outputs operational variables for each scenario along with investment variables for each time frame, $\overline{\triangle x}^{t}$ and $\overline{\triangle Q}^{t}$. 

\subsection{AC-PF feasibility}

The QP/LP step is followed by the AC-PF step, which is needed to maintain the AC PF feasibility destroyed by the linearization. Overall, combination of the QP/LP and AC-PF steps allow to maintain solution and  resolve contingencies of the system simultaneously and gracefully.

\section{Case studies}
\label{sec:case_studies}

The AC PF and optimization algorithms are implemented in Julia/JuMP. (See \cite{PowerModels} and references therein.) QP/LP optimizations (called at internal steps of our algorithm) are resolved by CPLEX \cite{CPLEX}. When possible we utilize IPOPT \cite{IPOPT}, called from JuMP, to solve the optimization problem (in its original, nonlinear formulation).  The (brute-force) IPOPT solution is computationally expansive and it is used as a ground truth (to validate our heuristic algorithm). Computational performance of the algorithms is analyzed on a Macbook Pro laptop (Core i7 3.3 GHz (2 Cores), 16 Gb of RAM).

\subsection{Algorithm validation}
\label{sec:validation}

The algorithm is validated by comparison with the IPOPT solution. Since IPOPT is not able to resolve the 2376 bus-large Polish model (even with a single scenario) we perform the initial validation study on the 30 bus IEEE model. (Both the Polish model and the IEEE 30 bus model are available within the MathPower package \cite{matpower}.) Table~\ref{tab2} presents results of the (IPOPT vs our heuristics) comparison,  
where the planning horizon is taken to be $1$ year, $T=5$, and $M=10$ for each time interval. Budget constraint per time interval is set to $\$ 200,000$. 

\begin{table}[h!]
\caption{Comparison of our heuristic algorithm against the (brute-force) IPOPT algorithm for the IEEE 30-bus model.} 
\renewcommand{\arraystretch}{2.5}
\centering 
\begin{tabu} to \columnwidth { | X[c] | X[c] | X[c] | X[c] | X[c] | X[c] |} 
\hline 
t. int.: & 1 & 2 & 3 & 4 & 5 \\  
\hline 
$\#$ sc.: & 10 & 10 & 10 & 10 & 10 \\
\hline 
alpha: & 1.02 & 1.04 & 1.06 & 1.08 & 1.15 \\
\hline 
dev.: & 0.001 & 0.003 & 0.005 & 0.01 & 0.02 \\
\hline   
\textbf{IPOPT} &  &  &  &  &  \\ 
\hline
SVC (bus 8), MVAR: & 4.00006 & 7.99995 & 10.5666 & 10.5666 & 10.5666 \\
\hline
SC (line 10), $\%$: & 0 & 0 & 1.661575 & 1.661618 & 1.661625 \\
\hline
\textbf{main} &  &  &  &  & \\
\hline 
SVC (bus 8), MVAR: & 4.0 & 8.0 & 10.5241 & 10.5241 & 10.5241 \\
\hline
SC (line 10), $\%$: & 0 & 0 & 2,321978 & 2,321978 & 2,321978 \\
\hline
\end{tabu}
\label{tab2} 
\end{table}

IPOPT objective is $6.029377e6$. Main algorithm objective is $6.029505e6$. Our main algorithm gives upper bounds solutions of the objective function with the gap less than 0.1$\%$.

\subsection{Scalability analysis}
\label{sec:scalability}

Our next step (after completion of the aforementioned validation study) was to perform a comparative analysis of the algorithms' (computational) scalability.  

\begin{figure}[h!]
\centering
\includegraphics[width=\columnwidth]{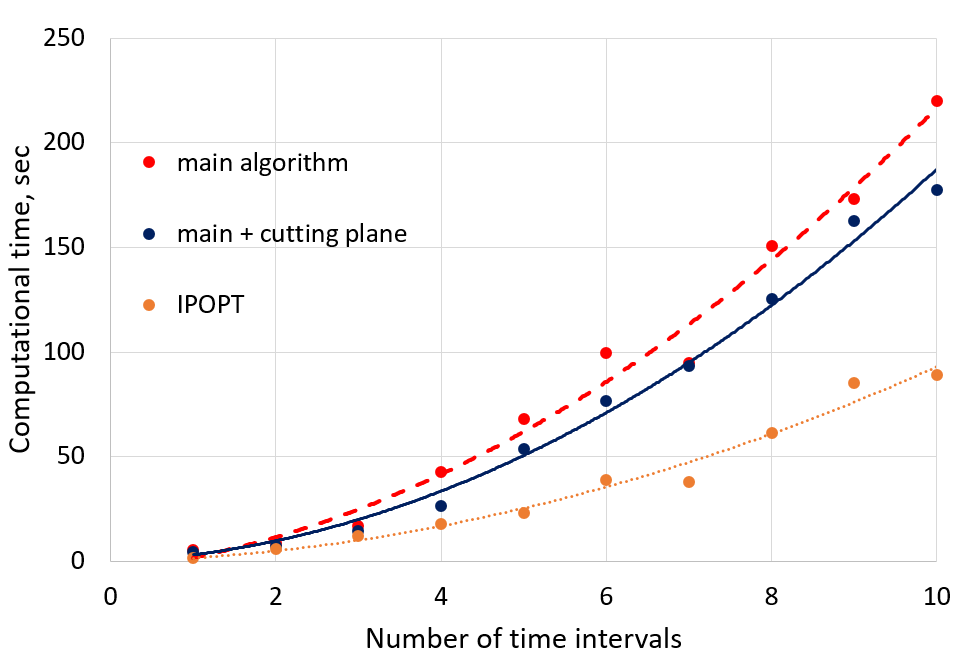}
\caption{Computational time of IPOPT (orange), of our main algorithm (red) and of our algorithm sped up with the cutting plane (dark blue) are shown as functions of the number of time intervals, $T$, for the 30 bus model. In all the tests shown the number of scenarios (per time interval) was $10$. Loading level is unity initially and it increases (imitating economic growth) by the factor $0.005$ per time step. Scenarios are generated with the deviation factor $0.01$. (See Appendix \ref{sec:scenarios} for details.)  Budget constraint (of $\$ 110,000$ per time step) is applied. Our algorithms (with and without cutting plane sped up) are limited to 20 iterations. 
\label{scalability_QP_DP}}
\end{figure}

\begin{figure}[h!]
\centering
\includegraphics[width=\columnwidth]{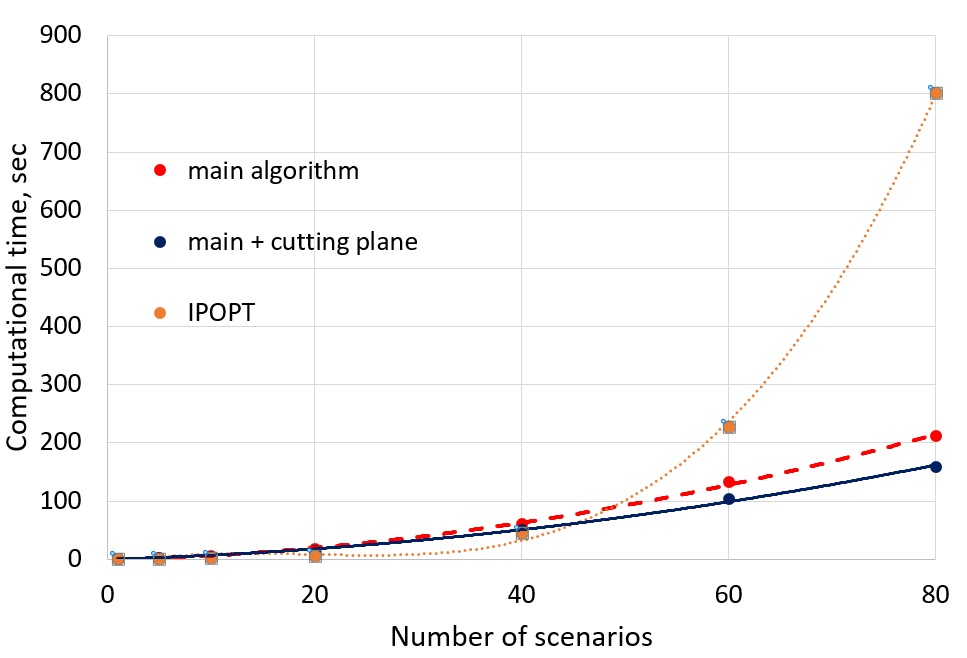}
\caption{Computational time of IPOPT (orange), of our main algorithm (red) and of our algorithm sped up with the cutting plane (dark blue) are shown as functions of the number of samples in the case of a single time interval for the 30 bus model. Loading level is set to $1.05$. Scenarios are generated with the deviation factor $0.01$. (See Appendix \ref{sec:scenarios} for details.)  No budget constraints are applied. Our algorithms (with and without cutting plane sped up) are limited to 20 iterations.
\label{scalability_QP_ScN}}
\end{figure}

First, we fix the number of scenarios ($10$ per time interval) and study dependence on the number of time intervals. The results are shown in Fig.~ \ref{scalability_QP_DP}. Then, we consider one time interval and study dependence on the number of scenarios. These results are shown in Fig.~ \ref{scalability_QP_ScN}. Both tests are done still on the 30 bus model with the quadratic generation cost (de-fault in the Mathpower package). In both cases we compare performance of the brute-force IPOPT solver with performance of our main algorithm and of the main algorithm reinforced by the cutting plane.



Comparing performance of the IPOPT in the two setting one observes a drastic difference. In the case of Fig.~\ref{scalability_QP_DP} IPOPT shows a surprisingly good performance, outperforming in speed both of our algorithms. 
(We relate this good performance of the IPOPT to 
using the primal simplex option within the IPOPT.) However the situation is reversed in the case of Fig.~\ref{scalability_QP_ScN} where, moreover, performance of the IPOPT degrades exponentially with the number of scenarios. Our main algorithm and the reinforced (by cutting plane) algorithm show similar scaling performance in both cases (with the reinforced algorithm performing slightly better). Notice that juxtaposing here our algorithms against each other has a sense because of the comparable number of constraints contributing the two optimization settings.


\begin{figure}[h!]
\centering
\includegraphics[width=\columnwidth]{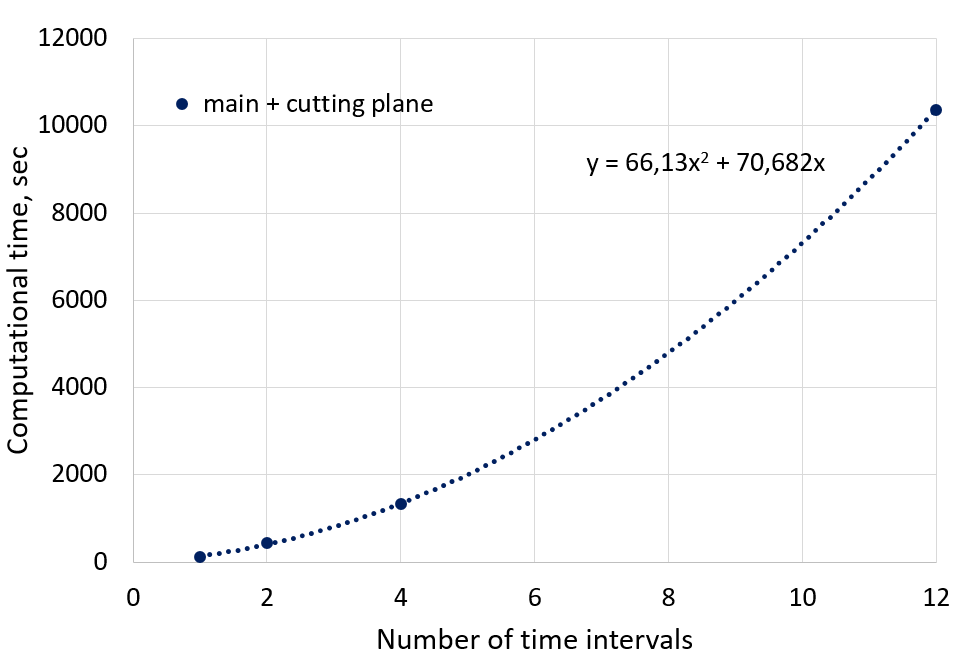}
\caption{Computational time of our main algorithm reinforced by the cutting plane shown vs the number of time intervals for the 2736 bus-large Polish model in the case of a single scenario (per time interval). The optimization cost is linear (according to the base case documented in Mathpower) and thus LP is used at each iteration step of our reinforced algorithm solved in $15$ iterations.  
In this case the loading level is set to unity in each time frame. 
Budget constraint of $\$ 50,000$ per time step is applied.
\label{scalability_LP_DP}}
\end{figure}

Moving to the scaling analysis of the Polish model, one first of all note that in this case the IPOPT fails to converge. To illustrate performance of our reinforced algorithm we focus on analyzing dependence on the number of time intervals. The results of our scaling experiments with the Polish model are shown in Fig.~(\ref{scalability_LP_DP}), where the dashed line show a (rather satisfactory) quadratic match (for dependence of the overall computational time on the number of intervals.) In this case we use the Primal Simplex CPLEX solver at each LP step of our algorithm. (This is LP and not QP, as in the 30 bus model, because the de-fault generation cost is linear in the Polish model of Mathpower.)  

We conclude this Section with a number of preliminary, and not yet fully conclusive but calling for further investigation, remarks. First of all, we have observed that developing efficient computational strategy for our linearization algorithm/heuristics became the task which is rather sensitive to the functional form of the generation cost and the choice of variables. If the cost is linear (in generated power) introducing auxiliary (line flow) variables and using the Primal Simplex solver is the winning strategy. However, the same approach in the case of quadratic cost (QP step replacing LP step) leads to slower convergence for the Primal Simplex algorithm while the barrier algorithm fails to converge at all. 




\subsection{Gradual investments to resolve congestion}
\label{sec:congestion}

\begin{figure}[h!]
\centering
\includegraphics[width=\columnwidth]{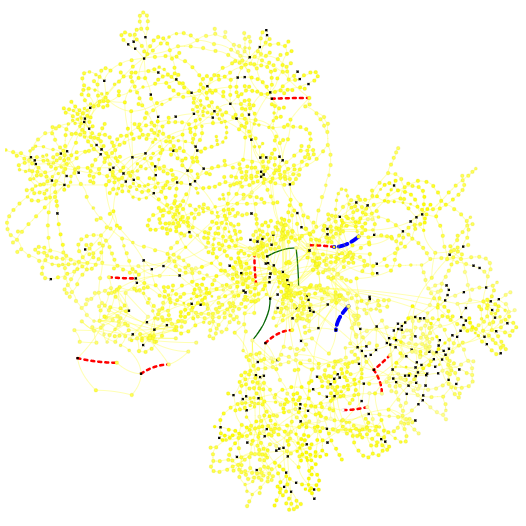}
\caption{Snapshot of the final solution (after all the investments are made). Red marks lines from the active set of cutting plane. Thin green marks lines with installed SCs which are not overloaded (initially). Dashed blue marks lines which are both in the active set (overloaded) and chosen for SC installation.
\label{polish1}}
\end{figure}

\begin{figure}[h!]
\centering
\includegraphics[width=\columnwidth]{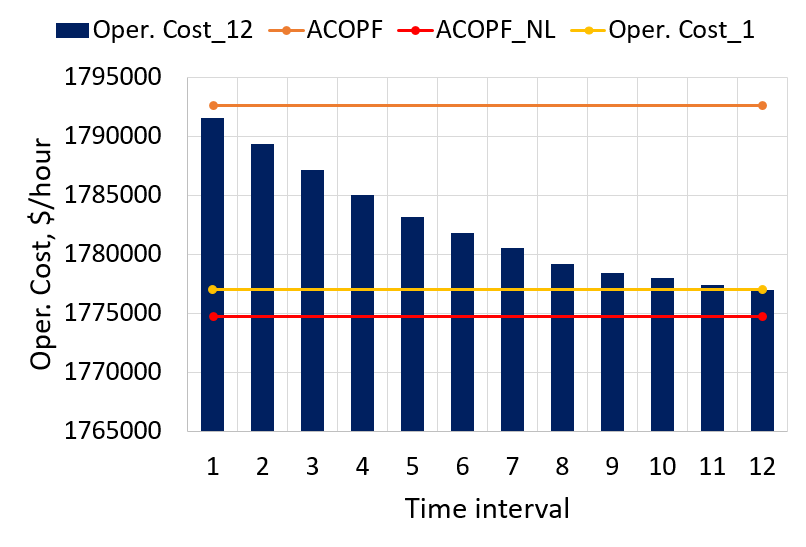}
\caption{Dependence of the optimal cost on  time (blue bars)  provided by our reinforced algorithm for the Polish model 
in the case of a single scenario (the same for different time intervals) and the year-long horizon split in $12$ periods (months). Orange line shows initial ACOPF cost (without investments). Red line shows ACOPF cost with thermal limits ignored. Yellow line corresponds to the optimal solution found in the case when the entire horizon is treated as a single time-interval.
\label{Polish1_oper}}
\end{figure}

\begin{figure}[h!]
\centering
\includegraphics[width=\columnwidth]{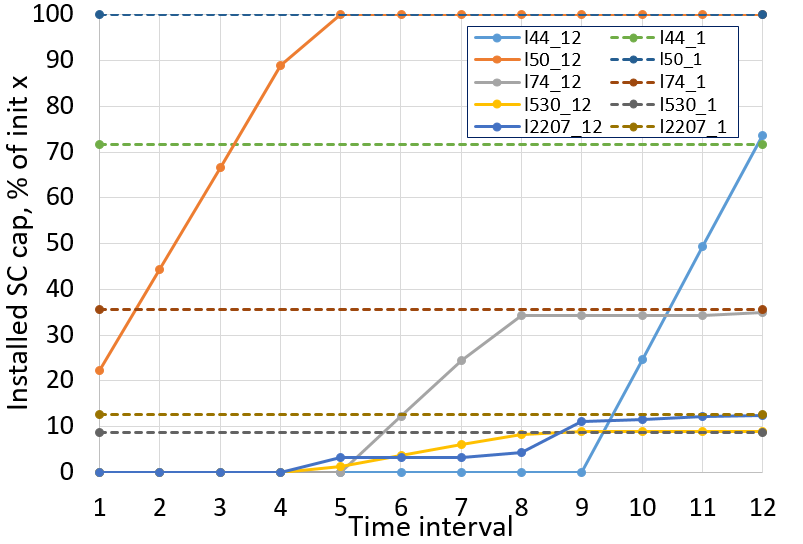}
\caption{Installed capacity of the SC devices at corresponding lines shown in percentage of the initial line inductances. This is the case of a single scenario therefore operational values coincide with respective capacities. For example, inductance of line $\#50$ is reduced to $0$.
\label{Polish1_SC_cap}}
\end{figure}

We apply our newly developed algorithm to study effect of the gradual investment, available only within the multi-time period framework, on the overall cost. We study the Polish model in the case of a single scenario with the optimization horizon of one year broken in $12$ periods. The (single) loading scenario, chosen to be stressed but still feasible (it is only $3\%$ away from the boundary of the AC OPF infeasibility - see Appendix \ref{sec:scenarios} for details), stays the same over time. The congestion cost of the initial loading scenario (yet no investments in FACTS)  is 17000 $\$/hour$. The investment budget is limited to $\$ 50,000$ per (one month) time interval. The results of optimal investment generated by our reinforced algorithm are illustrated in Figs.~\ref{polish1},\ref{Polish1_oper},\ref{Polish1_SC_cap}. We observe that only SC devices were installed at $5$ lines, of which only two would be overloaded (if the line limits are, first, ignored while solving AC OPF and then checked for the overload).  (Lines which are both overloaded, and thus contained in the active set of our cutting plance algorithm, and which are also selected for optimal SC installation are shown blue in Fig.~\ref{polish1}. Lines which are shown green were not overloaded but chosen for SC installation. Lines which are shown red were overloaded but were not chosen for SC installation.) 

We observe that the optimal installation is gradual. Moreover, all (constrained) available money are spent at each (time interval) decision. Fig.~\ref{Polish1_oper} shows how the congestion reduces with time, thus leading to reduction of the operational cost (blue bars) as time progresses. Orange line marks result of the AC OPF before investments start. Red line marks result for the (initial, i.e. before investments) AC OPF with the thermal limits ignored. 

Fig.~\ref{Polish1_SC_cap} shows that distribution of investments over lines and time period is nontrivial,  therefore utilizing the newly available SC-capacities with other operational degrees of freedom.

We also show in Fig.~\ref{Polish1_oper} and Fig.~\ref{Polish1_SC_cap} the result of optimal investment when the entire year (time horizon) is considered as one time interval. We observe that in the latter case the entire installation budget is used immediately such that the final solutions in the case one and $12$ intervals are the same and equal to the available budget. 
Notice, however,  that making one investment upfront for the entire year,  as opposed to breaking it into $12$,  periods is preferable because the overall (integrated over the year) cost of generation is reduced.  

\section{Conclusion}
\label{sec:conclusions}

In this manuscript, we propose a new optimization framework for algorithmic resolution of placement and sizing of FACTS devices in a large transmission grids. Our algorithmic solution of the problem takes into account non-linear power flow equations. The most important features of the newly developed algorithms include, \textbf{scalability}, allowing to resolve congestion over practical (thousands of buses) size transmission systems, and also the ability to \textbf{resolve multiple scenarios } and \textbf{over multiple time intervals simultaneously}. The optimization can be considered as generalizing standard AC-OPF over multiple scenarios and multiple time intervals with an added cost of installation. This optimization accounts for both \textbf{installation and operations}, thus allowing the installed devices to adjust operationally to a particular scenario within the bounds set by the installed capacity.

Many interesting cases were analyzed experimentally. In all the cases considered the output (optimal placement) is spatially \textbf{sparse} also showing strong \textbf{non-locality} (in the sense that placement of a device may resolve congestion in a distant region of the grid). It is also observed that under highly loaded conditions FACTS devices are beneficial in reducing the total cost of generation. Optimal installation of the devices helps to resolve infeasibilities that are projected to become even more severe in the future.

Main technical achievement reported in this paper is the development of an efficient heuristics for solving the non-linear,  non-convex and multi-time-interval optimization. The developed algorithm builds a convergent sequence of convex optimizations with linear constraints. Each constraint is represented explicitly through exact analytical linearization of the original nonlinear constraints (e.g. representing power flows and apparent power line limits) over all the degrees of freedom (including FACTS corrections) around the current operational point for particular loading scenario over particular time interval. In order to represent uncertainty in the projected growth of the system (loads) a custom scenario sampling split over multiple time intervals is introduced. Practicality of our approach for resolving the problem of investment (new installation) planning is illustrated on the IEEE 30-bus model and 2736-bus Polish model. It is evident from the experimental results that the approach is capable of both improving the system's economy (reduce congestion price and generation cost) and also of resolving feasibility issues by introducing additional degrees of freedom (associated with the newly installed FACTS devices). 

\section*{Acknowledgements}

The work at LANL was supported by funding from the U.S. Department of Energy’s Office of Electricity as part of the DOE Grid Modernization Initiative.

\appendices

\section{Transmission system modeling}
\label{sec:system_modeling}

\begin{figure}[h!]
\centering
\includegraphics[width=\columnwidth]{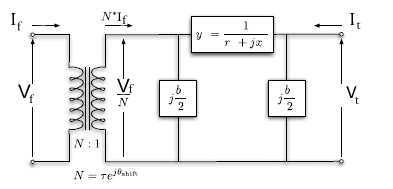}
\caption{$\pi$ model of a transmission line \cite{matpower}.
\label{trmod}}
\end{figure}

Figure \ref{trmod} shows a pictorial representation of $\pi$ model of a transmission line.

{\bf Power lines:}

The model is parametrized by series impedance $z = r + \myspecial{j} x$, total charging susceptance $b$, transformation ratio $\tau$  and shift angle $\theta_{\mbox{shift}}$. A transmission line has two ends: sending (often called \enquote{from} end) and receiving (often called \enquote{to} end).

Explicit expressions for the apparent powers injected at \enquote{from} and \enquote{to} ends of a line in terms of voltages and phases are:
\begin{eqnarray}
S_{f}(V_{f},\theta_{f},V_{t},\theta_{t},x)=\frac{V_{f}\left(r V_{f}- \tau V_{t}\left( r\cos\Delta+ x \sin\Delta\right)\right)}{\tau^2 (r^2+x^2)}\nonumber\\
-\myspecial{j} \frac{V_{f}}{2\tau^2 (r^2+x^2)}\Biggl(V_{f} (-2x+b (r^2+x^2))\nonumber\\
+2\tau V_{t}\left(x \cos\Delta+r \sin\Delta\right))\Biggr) \label{Sf}
\end{eqnarray}
\begin{eqnarray}
S_{t}(V_{f},\theta_{f},V_{t},\theta_{t},x)=\frac{V_{t} \left(r \tau V_{t}-V_{f} (r \cos\Delta+x \sin\Delta)\right)}{\tau^2 (r^2+x^2)}\nonumber \\
-\myspecial{j}\frac{V_{t}}{2\tau (r^2+x^2)}\Biggl(\tau V_{t} (-2x+b(r^2+x^2))\nonumber\\
+2v_{f} (x\cos\Delta-r\sin\Delta)\Biggr) \label{St}
\end{eqnarray}
where $\forall i:\quad v_i=V_i e^{{\it \bf j}\theta_i}$ and $\Delta = \theta_{f}-\theta_{t}-\theta_{\mbox{shift}}$.
Note that not only $S_{f}\neq S_{t}$, but also expressions for $S_{f}$ and $S_{t}$ are non-symmetric with respect to the change of $f$ and $t$ indices. The symmetry is broken as the transformer is positioned at \enquote{from} side of the line.

{\bf Generators and loads:}

Generators and loads are modeled as apparent power injections or consumptions:
\begin{eqnarray}
S_{gen} &=& P_{gen}+{\it \bf j}Q_{gen} \\
S_{load} &=& P_{load}+{\it \bf j}Q_{load}
\end{eqnarray}

Generation cost is a quadratic function of active power generation.

{\bf FACTS devices:}

FACTS devices are described by two variables - first is capacity, second is setting.

Series Compensation (SC) device capacity and setting for the line $\{i,j\}\in{\cal E}$:
\begin{eqnarray}
&\overline{\triangle x}_{ij}& \\
-\overline{\triangle x}_{ij} &\leq {\triangle x}_{ij} \leq& \overline{\triangle x}_{ij}
\end{eqnarray}

Static Var Compensation (SVC) device capacity and setting for the load node $i\in{\cal V}_l$:
\begin{eqnarray}
&\overline{\triangle Q}_{i}& \\
-\overline{\triangle Q}_{i} &\leq {\triangle Q}_{i} \leq& \overline{\triangle Q}_{i}
\end{eqnarray}

Expressions \eqref{Sf} and \eqref{St} are used for analytic linearization around current states.

\section{Generation of scenarios}
\label{sec:scenarios}

\begin{figure}[h!]
\centering
\def\svgwidth{0.6\columnwidth}
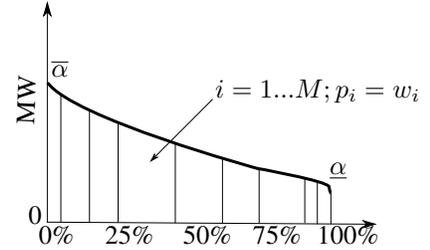
\caption{Piece-wise-constant approximation of the LD curve.
\label{fig:LDcurveAppr}}
\end{figure}

The method of scenario generation/sampling is used to include the uncertainty related to system load for the planning period.
Standard power system load growth over the time horizon is modeled via the LD curve. The base LD curve is illustrated in Figure~\ref{fig:LDcurveAppr}. In our simulations we usually have just a single base case loading profile for a power case. In order to define where it is situated on the corresponding LD curve we define loading level $\overline{\alpha}$ to represent condition which is $3\%$ from ACOPF infeasibility by uniform load rescaling (we call it top point conditions). 

We use the base LD curve, first, to generate LD curvess for consecutive years, re-scaling the base LD curve by the load growth factor of $0.5\%-1.5\%$ a year. Second, each early LD curve is split into $M$ piece-wise-constant parts. (We choose $M=6$ in our experiments.) Finally,  each piece of an LD curve is used to generate a scenarios according to a random (thus called sampling) procedure described below. This scheme of scenario generation/sampling models variations in the distribution of loads thus  simulating power system behavior during an extended period of time in the future.

We assume (and this assumption is maintained in all of our experimental tests) that each of the generated (sampled) load scenario is ACOPF feasible when the line constraints are ignored. (In other words, we consider the setting when there is enough of generation capacity even for the stressed cases.) Depending on the sampled scenario, 3 situations may arise.
\begin{enumerate}
\item ACOPF is feasible and congestion price is zero (low loading level).
\item ACOPF is feasible and congestion price is positive (higher loading level representing peak conditions).
\item ACOPF is infeasible due to either congestion of lines and/or voltage constraints but the system has enough generation capacity. ACOPF without apparent power limits on lines (and without voltage constraints if infeasible) is feasible (overloaded conditions which are possible in the future).
\end{enumerate}
The aim of planning installation of FACTS devices at the right locations with their corresponding capacities is to reduce generation cost for point 2, and to improve or extend feasibility domain of the system for the point 3. Extra years of service can hence be added to the existing grid by making it more flexible, thereby allowing to delay investments into new lines and generators.

\subsection{Scenarios sampling for each segment}
\label{subsec:scenarios_sampling}


The loading level $\alpha_i$ for a segment $i$ is represented by:
\begin{equation}
\alpha_i=\frac{\overline{\alpha}_i+\underline{\alpha}_i}{2}
\end{equation}
Future loading configurations are obtained from the base case by re-scaling all active and reactive loads by $\alpha_i$ uniformly. The resulting vector of loads for a segment is thus given by:
\begin{eqnarray}
    l_i^0=\alpha_i \times l^0
\end{eqnarray}
Loading configurations are generated, for each segment $i$ and each $j=1..N_i$, through modification of initial $l_i^0$. It is done by adding Gaussian correction to each load with zero expected value and a respective standard deviation:
\begin{eqnarray}
   & & l_i^j = l_i^0 + \mathcal{N}(0, \sigma_{l_i^0})  \label{Gauss}\\
   & & p_i^j = w_i/N_i \quad (\mbox{probability of a given scenario}) \label{Gauss-p}
\end{eqnarray}
where, $\sigma_{l_i^0}$ is given by:
 \begin{align}
  \sigma_{l_i^0}&=\frac{\overline{\alpha}_i-\alpha_i}{\alpha_i} \times l_i^0	\\
	& =\sigma \times l_i^0
 \end{align}
The choice of parameters used in our experimental test to sample the scenarios is described in Table~\ref{fig:ActualLDappr}.

\begin{table} [h!]
\caption{Implementation of the LD curve scheme} 
\centering 
\begin{tabu} to \columnwidth { | X[c] | X[c] | X[c] | X[c] |} 
\hline 
$i$ & $w_i$ & $\alpha_i$ & $\sigma$ \\ [0.5ex]  
\hline   
1 & 5,50 & 0,940 & 0,064 \\ 
2 & 19,50 & 0,845 & 0,041 \\
3 & 25,00 & 0,775 & 0,045 \\
4 & 25,00 & 0,685 & 0,080 \\
5 & 18,80 & 0,590 & 0,068 \\
6 & 6,20 & 0,51 & 0,078 \\  
\hline 
\end{tabu}
\label{fig:ActualLDappr} 
\end{table}

\subsection*{Congestion Analysis Correction}
\label{subsec:congestion}

If we study a case where for given load configuration standard AC-OPF outputs solution which is not congested,  i.e. solution for which each  constraint (on line flows or voltages) is satisfied with a  margin,  then this scenario does not require any FACTS installations. If the whole segment (from the procedure described in the preceding Subsection) is of this ``zero-congestion" type,  then obviously do not need to generate many samples representing the segment.  Instead,  we pick one re-scaled base scenario to represent the whole segment.

\section{Initializing the generation profile}	
\label{sec:init_gen}

Generation capacity is assumed to be large enough, i.e. not limiting, for the loading level considered.
The initial profile of the generation for each load scenario have to be determined to run the algorithm. The initial generation profile is derived following the following two steps.  1) Re-solving ACOPF with the thermal limits ignored. 2) Setting up proportional response of the generators. The second step includes a) search for the smallest load re-scaling factor $\alpha$ lowering the load and thus making the resulting case feasible; b) resolving ACOPF with this new re-scaled loading; c) increasing the generation and load proportionally to match the initial system loading (voltages are kept equal to these provided by the ACOPF solution); finally, d) solving ACPF to obtain generation matching the initial load.


%
%



%

\bibliographystyle{IEEEtran}
\bibliography{bibl_new}
%
%
%

\end{document}